# Towards the Safety-Relevant Dimension of Driver Behaviour: A Dual-State Model

Rulla Al-Haideri[1], Karim Ismail[2], Bilal Farooq[1], Adam Weiss[2]

[1]Laboratory of Innovation in Transportation (LiTrans), Toronto Metropolitan University, Toronto, M5B 1W7, Canada, [2]Department of Civil and Environmental Engineering, Carleton University, 1125 Colonel By Dr, Ottawa, ON K1S 5B6, Canada

**Abstract**

Improving road safety requires more than identifying collisions or near-misses; it calls for a deeper behavioural understanding of how drivers perceive and respond to risk in real time. Traditional approaches to modelling driver behaviour have often relied on simplified assumptions, such as treating behaviour as homogeneous across drivers or as independent of context. This paper makes a methodological contribution by introducing a new dimension of traffic conflict severity: the probability that a driver is in a defensive state. This behavioural probability reflects an internal response to perceived risk and is estimated using a latent class Discrete Choice Model (DCM) that captures driver behaviour as a probabilistic mixture of two latent driving states: a *defensive* state, representing heightened caution and collision-avoidance intentions under perceived risk, and a *neutral* state, reflecting routine driving behaviour under low-threat conditions. The proposed framework does not initially rely on collision data, offering the potential to assess safety proactively using only naturalistic trajectory patterns. The framework is grounded in psychological theory, particularly the triad of affect, behaviour, and cognition. It is also informed by two key concepts. First, that event severity exists on a continuum, rather than being confined to binary categories of safe or unsafe. Second, that drivers perceive risk through a dynamic spatial safety field, one that varies with direction, proximity, and the motion of surrounding road users. These concepts support a behaviourally-grounded latent class DCM in which driving states are inferred from trajectory-derived features, and transitions between them are represented probabilistically, rather than as binary shifts. Applied to the publicly available rounD dataset, the framework yields interpretable estimates of state membership probabilities. The defensive state consistently reflects stronger sensitivity to spatial and temporal risk, while the neutral state captures context-appropriate yet less reactive driving patterns. Importantly, the paper also proposes a method to assess the quality of the estimated probability of being in a defensive state. Because of the duality between the defensive and neutral states, evaluating the consistency of one offer insights into the reliability of the other. To explore this, a multi-step validation procedure is applied across five data subsets representing different driving contexts, including free-flow and diverging scenarios, to examine how well the neutral state generalises beyond the estimation sample. The results indicate that the estimates may be behaviourally consistent and appear reasonably stable across contexts, suggesting that the defensive probability could serve as a useful behavioural dimension of traffic conflict severity.

## 1. Introduction

Understanding the complexity of driver behaviour remains an important step toward addressing ongoing concerns in road safety. It is widely recognised that human factors contribute to the majority of traffic incidents, and much of the literature suggests that their role may be substantial in explaining why crashes occur (US Department of Transportation, 2004). At the same time, existing driver-behaviour models have typically depended on oversimplified assumptions, such as, treating behaviour as uniform across all drivers and ignoring the influence of contextual factors. However, driver decisions are frequently shaped by a combination of internal states and external conditions, and it is becoming increasingly important to account for these behavioural nuances when exploring the causes and correlates of unsafe outcomes (Jeon et al., 2014; Ma et al., 2023). Unsafe actions such as speeding, distraction, and misjudging traffic conditions are frequently linked to lapses in situational awareness or perceptual errors. These may arise from occlusions, divided attention, or incorrect mental models of surrounding traffic (Kircher and Ahlstrom, 2017). Drivers must continuously maintain an accurate and dynamic understanding of their environment, especially under time pressure or in unexpected situations (Endsley, 1995; Schaap, 2012). Understanding these behavioural mechanisms is not only theoretically important, but foundational for improving safety outcomes.

This paper conceptualises traffic severity as a multi-dimensional phenomenon rather than a binary classification of safe versus unsafe outcomes. This conceptualisation is in line with Zheng et al. (2021), who identify three of these dimensions of traffic conflict severity: (1) the consequence of a potential collision, referring to expected injury or damage; (2) the proximity between road users in time or space; and (3) the evasive action made by the road user. Building on this framework, this paper introduces a new dimension of conflict severity, the driver's behavioural response to perceived risk. Unlike the observable measures noted above, this dimension captures an internal cognitive-affective state and is represented through a probabilistic latent class structure.

Specifically, reflecting the dual-process theories of cognition (Kahneman, 2011) and the blended nature of human emotion (Larsen et al., 2001), as will be elaborated in Section 2. Each driver at a given moment is conceptualised as being in a probabilistic mixture of two latent behavioural states: a defensive state and a neutral state. This continuous and complementary probability of membership directly reflects the dynamic interaction and trade-off in behavioural tendencies under varying levels of perceived risk, moving beyond binary classifications towards a more nuanced, psychologically-informed representation of driver state. The probability of being in the defensive state is considered safety-relevant, as it reflects a collision-avoidance orientation in response to perceived risk. This probability serves as the operational measure of the proposed behavioural dimension. The conceptualisation aligns with psychological insights, discussed in Section 2, that human responses often involve a blend of intuitive (System 1) and deliberative (System 2) processes, and that emotional states are rarely singular but rather a dynamic blend of underlying dimensions. As such, it is proposed as a behavioural conflict severity dimension that complements the existing dimensions of severity.

This behavioural dimension can be conceptually related to proximity and could, in fact, be viewed as a latent manifestation of it. In this sense, what is modelled as a behavioural state might reflect an interpretation of external proximity cues, a process likely involving the blended cognitive and affective mechanisms detailed in Section 2. We do not claim that this dimension is entirely unique or distinct to the existing dimensions, particularly proximity. Rather, it is proposed as a promising extension that warrants further exploration. The behavioural framework is flexible enough to account for such potential interdependencies, allowing behavioural response and proximity to coexist and interact without assuming complete independence. Importantly, these multiple dimensions of traffic conflict severity do not always align in predictable or consistent ways. For instance, a driver may engage in defensive behaviour even when no immediate threat is present or may face severe potential consequences without being in close proximity to a collision. Similarly, indicators of proximity may not correspond directly to the inferred behavioural state, and the success of evasive actions may depend as much on environmental constraints as on driver intent. It is assumed that most road users generally act in rational and adaptive ways in response to perceived risk; however, the behavioural and situational landscape in which these decisions unfold is inherently dynamic and shaped by varying contextual configurations.

Following the ideas of Svensson and Hydén (2006), who conceptualise conflict severity as a continuum, the proposed approach avoids strict classifications of latent behavioural states. Instead, it estimates the likelihood of defensive or neutral behaviour to reflect the evolving nature of road user responses. This flexible framing accounts for blurred boundaries between routine and cautious driving. This perspective is further supported by Laureshyn et al. 2010), who argue that severity unfolds over time through interaction, rather than being captured at a single point via measures such as Time-to-Collision (TTC). In this sense, trajectories are not just position data; they may represent behavioural traces of how users interpret and respond to dynamic surroundings. The risk-field concept, proposed by Sahu et al. (2025) is adopted to contextualise driver behaviour spatially. This concept treats road users as being surrounded by directional, dynamic fields of risk. Analogous to electric fields, these vary in intensity and direction, with greater sensitivity often observed in the forward direction. The concept is operationalised by modelling a spatial grid around each driver, shaped by their speed and heading, which may help identify when and where defensive responses are more likely to emerge.

To consolidate the ideas introduced above, Figure 1 presents a conceptual framework that links trajectory-derived variables with a latent behavioural model of defensive driving. The framework illustrates how features such as speed, proximity, and conflict indicators feed into a probabilistic model that estimates drivers' behavioural state as a continuum between defensive and neutral modes. This behavioural estimate reflects an internal response to perceived risk and is treated as a complementary dimension alongside existing severity measures. By incorporating core theoretical perspectives, including the ABC triad (Affect, Behaviour, Cognition) and the concept of conflict severity as a continuum, the framework helps ground our modelling approach and clarifies the behavioural contribution of the model within a broader safety context.

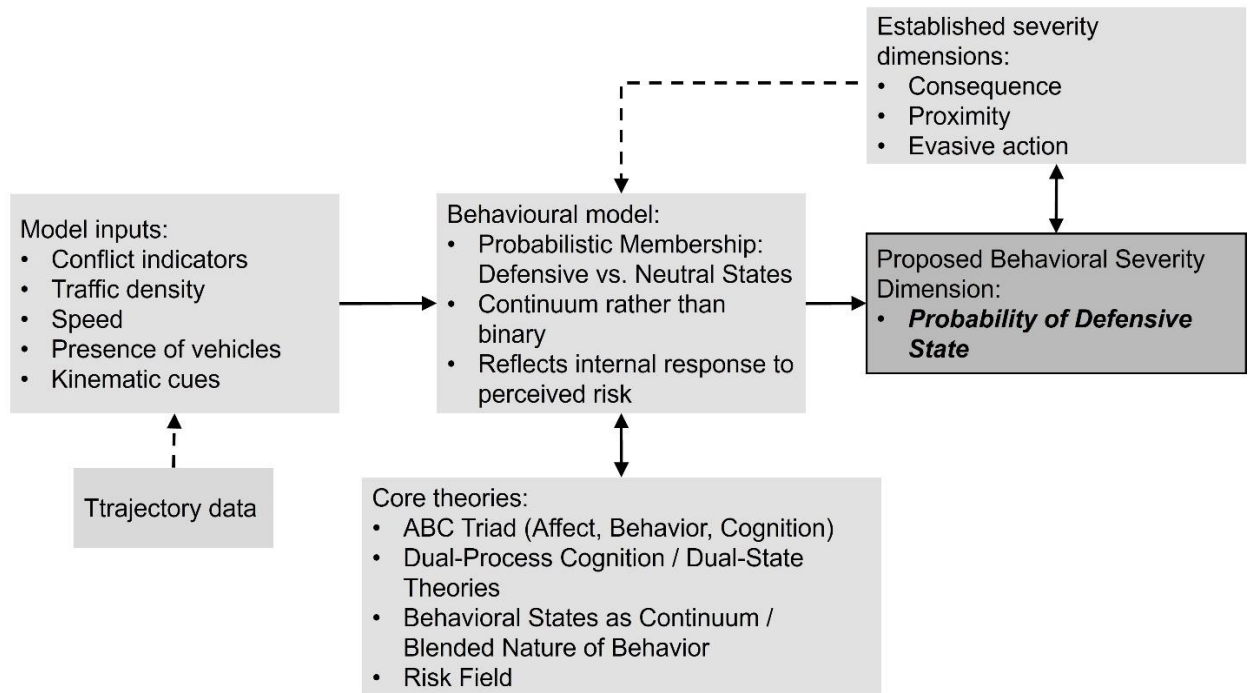

*Figure 1 Conceptual framework for obtaining the newly proposed severity dimension: probability defensive state*

The ideas presented here form the basis for a novel framework that uniquely blends established behavioural theory, empirical conflict indicators, and trajectory data. This paper contributes to the literature by proposing an approach to estimate the behavioural response dimension, which is often left implicit and assumed to reside in the mindset of the road user. We introduce a probabilistic framework, aimed at estimating the continuous likelihood that a road user is in a defensive state. This estimation is based on their dynamic responses to spatial and temporal conditions, providing a direct behavioural interpretation that complements traditional proximity-based indicators. In the sections that follow, we aim to develop and apply this framework to answer the following research questions:

1. Does the proposed dual-state latent model offer better explanatory power than a single-state model?
2. How can driver behaviour be modelled as a probabilistic mixture of latent behavioural states using trajectory data?
3. How can we specify the latent class membership and conditional choice components to reflect cognitive and operational layers of driver decision-making?
4. How can we assess the quality and interpretability of the model, particularly in terms of validating that the defensive and neutral states align with expected behavioural patterns and risk conditions?

The paper is organised as follows. Section 2 presents the theoretical foundation for the blended and dynamic nature of human behaviour. Section 3 introduces the proposed methodological framework. Section 4 applies the model to a naturalistic trajectory dataset, reporting estimation results, behavioural interpretation, and validation of the neutral state. Section 5 concludes the paper by discussing the implications of defensive-state probability as a behavioural safety indicator and outlining directions for future research.

## 2. Theoretical Foundations: The Blended Nature of Human Behaviour

Building on the behavioural perspective outlined in the introduction, this section draws on foundational psychological theories to explore how internal states, such as emotion, cognition, and action, interact in shaping road user behaviour. The core idea here is that behaviour, particularly in dynamic contexts like driving, is rarely dictated by a single factor. Instead, it emerges from a blend of simultaneous influences, each contributing in varying degrees.

In psychology, human action is often conceptualised through three interacting components: affect, behaviour, and cognition (Breckler, 1984; La Guardia et al., 2000; van Harreveld et al., 2015). Commonly known as the ABCs of psychology, these elements reflect how individuals experience, evaluate, and respond to their environment. Affect refers to emotional states, feelings and moods that colour how we interpret risk. Behaviour refers to observable responses, from reflexive reactions to deliberate actions. Cognition captures how individuals process information, including their beliefs, perceptions, and expectations about road situations (Shouse, 2005; Greeno et al., 1996). These components are deeply influence each other. For example, how a road user interprets a rapidly approaching vehicle (cognition) may influence their stress response (affect), which in turn may shape whether they brake or accelerate (behaviour). These feedback loops are central to behavioural models such as Cognitive Behavioural Therapy, which explicitly address how thoughts, feelings, and actions influence one another (Bandura, 1978; Beck, 2021). Understanding this interaction is critical because our goal is to identify when a driver transitions into a more defensive state, a shift that may stem from cognitive recognition of risk, an affective sense of discomfort, or both. By modelling this response as a probabilistic outcome, we can move beyond surface-level behaviours and attempt to infer the internal states driving them. The next two subsections elaborate on how blending operates within two key domains: emotion and cognition.

## 2.1. The Blended Nature of Human Emotion

Traditional views often conceive of emotions as discrete, independent entities. However, a growing body of research, often referred to as the study of mixed emotions, blended emotions, emotional complexity, or the dimensional approach to emotion, challenges this notion. Instead, these theories propose that our emotional experiences are often rich combinations of more basic emotional components or can be described along continuous dimensions rather than as distinct, separate categories.

A foundational contribution to this understanding comes from Russell (1980), who proposed a circumplex model of affect. In this model, emotions are organized around a circle defined by two core dimensions: valence (ranging from pleasure to displeasure) and arousal (ranging from activation to deactivation). Within this 2D space, any specific emotion is represented as a point, implying that emotions are not isolated categories but rather dynamic blends of these underlying dimensions. For instance, "anger" might be characterized by high arousal and negative valence, while "calmness" would be described by low arousal and positive valence. Building on such dimensional approaches, Watson and Tellegen (1985) further solidified the prominence of positive and negative affect as major components of mood, suggesting that even seemingly simple moods are, in fact, composites.

Empirical evidence further supports the notion of blended emotions. Larsen et al. (2001) provided compelling evidence that individuals can indeed experience mixed emotions simultaneously, such as feeling both happy and sad at the same time. This finding directly challenges the idea of purely discrete and mutually exclusive emotional states, demonstrating that positive and negative affect can co-occur. While broader in scope, research like that by Ong et al. (2006) on psychological resilience implicitly acknowledges that individuals coping with stress often experience a complex mixture of emotions, including both positive and negative feelings, as part of their adaptive mechanisms. Furthermore, Zajonc (1980) work, while primarily focusing on the primacy of affect, frequently touches upon the interwoven and complex nature of emotional responses, reinforcing the idea that emotional experiences are rarely "pure" and are instead influenced by various interacting factors, leading to a blended overall experience.

## 2.2. The Blended Nature of Human Cognitive Processes

Similar to emotions, human cognitive processes are also increasingly understood as a mixture of different modes and pathways rather than a singular, unified system. This concept is frequently explored through dual-process theories, parallel processing, integrated cognitive models, and the interplay of different cognitive systems. These theories propose that cognition involves the simultaneous or interactive operation of multiple, sometimes distinct, processing pathways, rather than a single, linear operation. The most influential framework in this area is the dual-process theory, famously popularized by Kahneman (2011) in his work Thinking, Fast and Slow. This theory distinguishes between System 1 (fast, automatic, intuitive, and heuristic-based thinking) and System 2 (slow, effortful, deliberate, and rule-based thinking). The fundamental insight is that both systems operate in parallel and interact dynamically. System 1 often generates initial responses, which System 2 may then monitor, correct, or override. Consequently, our cognitive output is a direct result of the continuous interplay and blending of these two distinct modes of thought.

Further emphasising this relationship, Stanovich and West (2000) extensively discuss how individual differences in these System 1 and System 2 processes contribute to variations in reasoning and decision-making. Their work highlights that actual cognitive performance arises from the dynamic interaction between these two distinct types of processing. A comprehensive review by Evans (2008) on dual-processing accounts across various cognitive domains underscores the widespread acceptance of the idea that cognitive processes inherently involve a mixture of at least two distinct types of processing. Beyond dual-process theories, Parallel Distributed Processing (PDP) models, notably advanced by Rumelhart et al. (1986), offer another compelling perspective. These models propose that cognitive processes, such as perception, memory, and language, emerge from the simultaneous activation and interaction of many simple processing units (analogous to neurons). This framework strongly argues for cognition being a "mixture" of parallel, distributed activities rather than sequential, modular operations. For example, McClelland and Davd (1981) interactive activation model of context effects in letter perception exemplifies how different levels of processing (*e.g.*, features, letters, words) interact in parallel to produce a final cognitive output, thereby illustrating a "mixture" of influences at different processing stages.

In summary, this section demonstrates that human behaviour, particularly in dynamic and high-stakes environments like driving, is inherently a complex interplay of blended emotions and dual cognitive processes. The evidence presented from emotional circumplex models and dual-process theories firmly establishes that our internal states are rarely singular or purely discrete, but rather continuous mixtures and dynamic interactions. This foundational understanding directly informs our methodological approach, where the driver's state is not conceptualised as a fixed, categorical attribute but as a probabilistic continuum between defensive and neutral poles. The method presented in Section 3 is designed to capture this inherent duality and blended nature by estimating the varying degrees of these latent states based on observable driving patterns and contextual cues.

## 3. Methods

We employ a Discrete Choice Modelling (DCM) framework to analyse driver behaviour from vehicle trajectory data. The model is structured around a utility function that includes both a systematic component, reflecting observable variables such as speeding, turning, or conflict indicators, and a random component, which captures unobserved variability in driver decision-making. Within this framework, we first construct a spatial choice set from the observed trajectories to enable the subsequent application of a dual-state latent class extension designed to capture variations in drivers' risk attitudes. The objective of this method is to identify and interpret latent driver behavioural states, defensive and neutral, based on trajectory data, and to validate the robustness of these states across different driving contexts. To fully contextualise this modelling strategy, we first delineate its core theoretical foundations.

### *3.1. Theoretical Framework*

This work adopts a dual-state view of driver behaviour, where driving is understood as a probabilistic mixture of two latent behavioural states. One state reflects a defensive mode, where the driver is more preoccupied with collision avoidance. The other state represents a more neutral state, where the driver is focused on routine path-following and progression toward a destination. Below, we outline several theoretical elements that guide our modelling approach:

**Element 1: Driving behaviour is blended, not binary.**

Drivers may operate with varying degrees of defensiveness or neutrality, depending on situational context. Rather than switching discretely between states, their behaviour is assumed to reflect a continuous blend between the two. This perspective aligns with the view that behavioural responses to risk unfold along a continuum, not as either-or events.

**Element 2: The defensive state is introduced as a safety-relevant dimension.**

The degree of membership in the defensive state may reflect a safety-relevant behavioural marker, offering insight into how drivers perceive and respond to potential threats. While this measure is not intended to directly predict crash risk, it provides a way to observe safety-conscious behaviour in naturalistic data. The extent to which this dimension aligns with actual crash outcomes remains an open question for future research.

**Element 3: Membership in the defensive state captures perceived severity.**

The model represents the severity of a perceived driving situation through the probability of defensive state membership. A higher probability indicates stronger behavioural signs of avoidance or caution. Rather than relying on binary indicators such as near-misses or collisions, this approach enables a more gradual understanding of risk.

**Element 4: Free-flow driving serves as a practical reference for the neutral state.**

A key challenge in modelling defensive behaviour lies in the absence of a clear benchmark for what constitutes a purely defensive state, such behaviour is difficult to capture and rarely documented in scientific datasets. In light of this, we turn to the duality between the defensive and neutral states to guide model interpretation. The neutral state, in contrast, can be approximated more readily under free-flow conditions, where drivers are not actively interacting with others and are likely focused on maintaining speed and following their intended path. While free-flow conditions are not entirely risk-free, single-vehicle incidents or roadside hazards may still occur, they offer a reasonable and observable baseline for routine, lower-risk driving. Evaluating how well the model captures this neutral behaviour provides insight into its ability to distinguish the more elusive defensive state.

**Element 5: Model validity is interpreted through the duality of the states.**

Since there is no definitive benchmark for what constitutes a "pure" defensive state, particularly in naturalistic datasets, model interpretation relies on the complementarity between the two states. If neutral behaviour can be reliably identified under free-flow conditions, this provides indirect support for the model's ability to isolate defensive states in more complex settings. In this way, model validity is approached not through direct measurement of defensiveness, but by assessing how well its complement (neutrality) aligns with observable patterns.

**Element 6: Contextual conditions are expected to influence state membership.**

The likelihood of being in either state is assumed to vary with external factors, such as traffic density, relative speed, or the presence of surrounding vehicles. These contextual variables serve not only as model inputs but also as interpretable drivers that shape the behavioural state distribution.

## 3.2. Data, Choice-Set Construction and Observed Choices Extraction

The basis of the analysis is a dataset of vehicle trajectories, where each observation provides a vehicle's two-dimensional position $(x, y)$, speed, and acceleration over time. At each time step, the set of feasible next positions is discretized into a polar-coordinate grid composed of speed rings and turning cones (see Figure 2). The driver's actual observed choice is identified by calculating the radial change in speed $(\Delta r)$ and angular change in direction $(\Delta \theta)$ between successive time steps. Each $(\Delta r, \Delta \theta)$ pair is then mapped to its corresponding cell in the spatial choice set grid. Full details on grid resolution, regime boundaries, and choice extraction can be found in Al-Haidari et al. (2025).

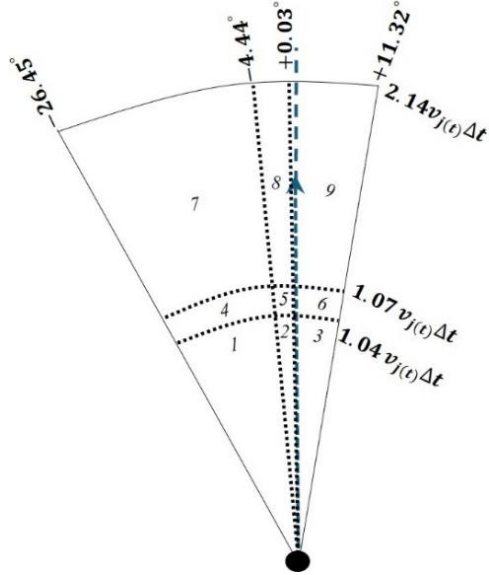

*Figure 2 Spatial choice set grid*

Following the construction of the choice set grid, observed choices are first extracted for the entire dataset. These extracted choices serve as the foundation for model estimation and validation. A method is then proposed to validate the characteristics of the neutral state identified by the dual-state latent DCM, which we hereafter refer to as the dual-state model. This validation is essential to ensure that the neutral state captured by the model aligns with expected neutral driving behaviour in different driving contexts. Figure 3 provides a schematic overview of this analytical process, illustrating the flow from raw trajectory data to the final validation step. The processed dataset is segmented into three distinct subsets based on the presence of vehicle interaction and the calculation of a conflict indicator:

(1) Interactions Data: data where a specific conflict indicator is calculated, corresponding to scenarios where the driver is interacting with other vehicles in a manner that meets the indicator's criteria;
(2) Free-Flow Data: data with no interaction with other vehicles, representing free-flow or isolated driving conditions; and
(3) Diverging Interactions Data (no indicator): data where no conflict indicator is calculated, but the driver is still interacting with other vehicles, specifically in scenarios where the interacting vehicle is diverging or moving away.

These subsets span a continuum of interaction intensity, from complex, conflict-rich scenarios to uninhibited free-flow, allowing validation of whether the neutral state behaves consistently across low-risk conditions. Importantly, only the Interactions data is used to estimate the dual-state DCM. The free-flow and diverging interactions data serve as hold-out subsets for validation purposes, allowing assessment of whether the neutral state behaves consistently across lower-risk conditions.

To validate the neutral state component identified by the dual-state model, standalone models are estimated using the free-flow and diverging interactions data. This approach is adopted because these three data subsets represent different driving contexts, ranging from complex interactions to purely free flow driving conditions. Rather than directly comparing parameter estimates across these models, which is not meaningful due to differing data contexts and utility scales, a predictive validation strategy is proposed. Specifically, the predictive accuracy of the neutral state component is evaluated by comparing it to the independently estimated models from the hold-out subsets.

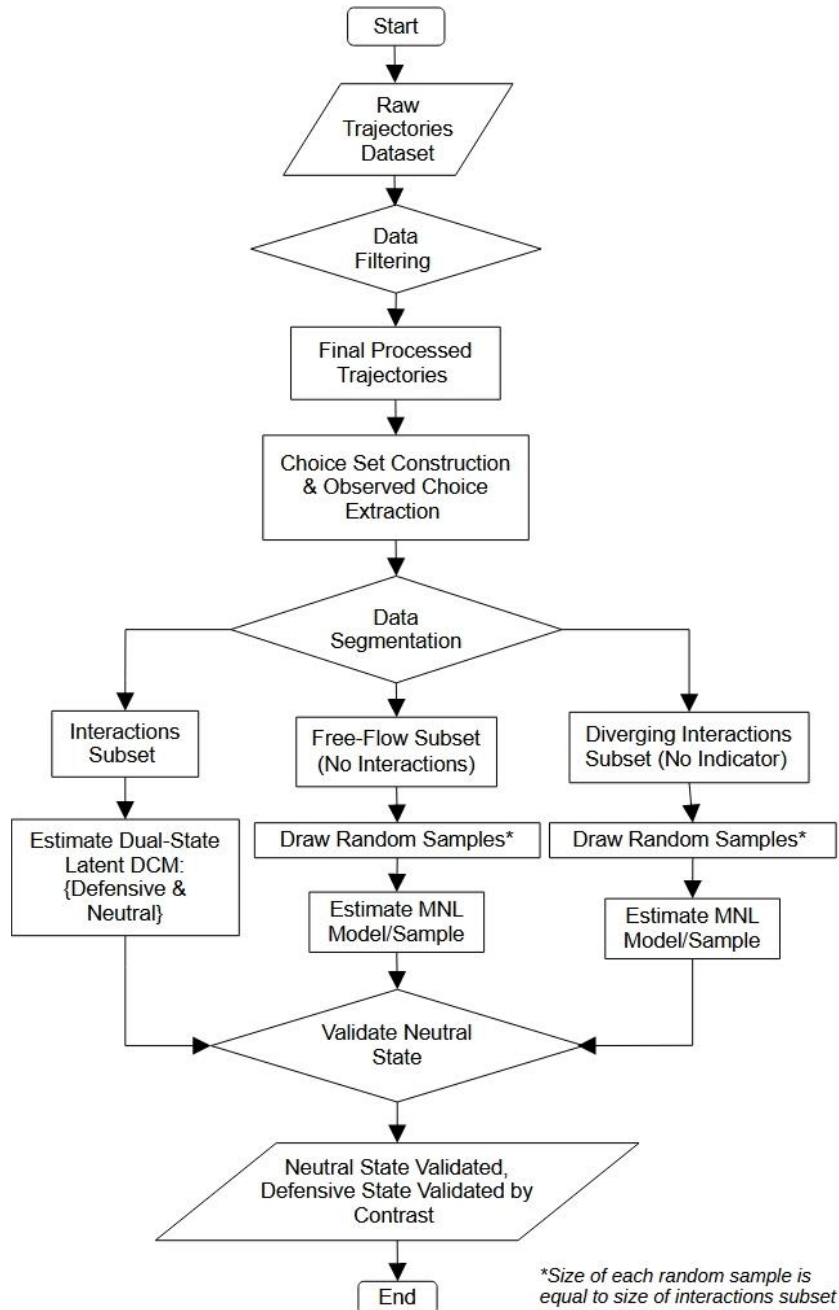

*Figure 3 Schematic overview of the proposed methodology for defining and validating latent states*

*3.3. Model Specification and Estimation Procedure*

To specify a model that accounts for the inherent variability in drivers' risk attitudes, the DCM framework is combined with a two-class latent structure. This framework assumes that, at any given time, a driver's behaviour is best represented as a probabilistic mixture of two latent behavioural states: a defensive state, in which the primary focus is on collision avoidance, and a neutral (or normal) state, where the driver focuses on efficient, uninterrupted travel.

In latent class models, individuals are assumed to be implicitly segmented into a finite number of unobserved behavioural classes (denoted by $S$), though their actual class membership is not directly observed. Each class is associated with its own set of behavioural rules or utility parameters. The probability of individual $n$ selecting alternative $i$ from a choice set of $I$ alternatives at time $t$ is obtained by aggregating the class-specific choice probabilities, weighted by the probability of the individual belonging to each class. This is expressed as (Greene and Hensher, 2003):

$$P_{nit} = \sum_{s=1}^{S} P_{nt(s)} \cdot P_{nit}|(s)$$

Where:

- $P_{nt(s)}$ is the probability that individual $n$ belongs to class $s$ at time $t$ (class membership probability); and
- $P_{nit}|(s)$ is the probability that individual $n$ selects alternative $i$ at time $t$, conditional on being in class $s$ (conditional choice probability).

This formulation captures unobserved heterogeneity by allowing each class to exhibit distinct sensitivities to explanatory variables. In this framework, the class membership probability identifies which behavioural state a driver is likely to adopt, while the conditional choice probability defines the movement decisions within that state. To ensure conceptual clarity, the variables used in each probability must be selected based on their behavioural function.

The class membership probability should include variables that capture a driver's general situational awareness or perceived environmental risk, such as traffic volume, density, or sustained closing patterns. If a conflict indicator is used here, it should reflect an abstract or anticipatory assessment of risk rather than an immediate threat.

By contrast, the conditional choice probability should rely on temporally and/or spatially precise indicators that reflect immediate real-time collision threats, such as relative speed, heading, or proximity to surrounding vehicles. These risk indicators are active only in the defensive state, where they influence operational decisions such as braking or evasive manoeuvres. In the neutral state, drivers operate under low-risk conditions focus on stable, goal-directed behaviour. Utility in this state excludes collision-related terms and emphasises momentum preservation, trajectory smoothness, and destination-oriented movement.

A key feature of our model is that the defensive class could be explicitly defined through interpretable behavioural terms, such as deceleration or response to emerging threats. This prevents the model from inferring class structures in arbitrary or non-behavioural ways. Because risk indicators (*e.g.*, TTC) have physical meaning, the model's behavioural assignment can be evaluated for plausibility. If defensive responses are associated with low-risk conditions, this may indicate a misspecification rather than a valid latent segmentation. Embedding interpretability in the model supports both diagnostic transparency and behavioural validity.

Although the framework is implemented using a Multinomial Logit (MNL) specification for the class membership and conditional choice probabilities in the application presented in this paper, it is compatible with other DCM structures. Depending on the behavioural context and data characteristics, models such as generalized extreme value or mixed logit may be adopted to capture more complex decision processes while still leveraging the latent class mechanism to account for unobserved heterogeneity.

Model estimation is performed using the Expectation-Maximization (EM) in combination with Maximum Likelihood Estimation (MLE) (Bhat, 1997; Kim and Mokhtarian, 2023). The EM algorithm iteratively alternates between estimating class membership probabilities and updating the conditional choice model parameters. This process continues until convergence, defined either by a negligible change in log-likelihood or when the Root Mean Square Error (RMSE) of parameter estimates across iterations falls below 5%. This convergence threshold ensures numerical stability and reliable inference.

### *3.4. Validation of the Neutral State*

To assess the behavioural consistency and predictive accuracy of the neutral class identified by the dual-state model, a multi-layered validation strategy is implemented across several driving contexts (data subsets). This validation serves two critical purposes. First, it evaluates the generalizability of the neutral class parameters, testing whether they accurately reflect baseline, non-defensive driving behaviour beyond the dataset used for model estimation. Second, it ensures that the latent behavioural segmentation is meaningful, rather than merely overfitted to the estimation sample. Importantly, the purity of the neutral class extraction directly affects the reliability of the defensive class coefficients. The clearer and more stable the definition of neutral behaviour, the better the model can isolate and capture the behavioural adaptations that characterise the defensive state. The predictive accuracy and behavioural consistency of the neutral class is evaluated using five data subsets, each representing a distinct driving context:

1. *Interactions Subset – Neutral Class Application*:

   This subset is identical to the one used for estimating the dual-state model. It serves as a best-case benchmark, as the neutral class is calibrated on this data. The neutral class parameters are reapplied here to confirm internal consistency and baseline predictive performance.

2. *Free-Flow Subset – Independent MNL Model* Estimation:

   From the free-flow driving data, 20 random subsets are drawn, each containing the same number of observations as the interactions dataset. The first 10 subsets are used to estimate 10 standalone MNL models specified with the same utility structure as the neutral class. This enables an independent benchmark under routine, unconstrained driving conditions.

3. *Free-Flow Subset – Neutral Class Application*:

   The remaining 10 free-flow samples are used to apply the neutral class parameters from the latent class model to predict choices. This provides insight into how well the neutral class generalises to free-flow conditions that were not used during model estimation.

4. *Diverging Interaction Subset – Independent MNL Model Estimation*:

   Similarly, 20 random samples are drawn from diverging interaction subset, where vehicles are moving away from the driver. The first 10 samples are used to estimate 10 standalone MNL models specified with the same utility structure as the neutral class.

5. *Diverging Interaction Subset – Neutral Class* Application *Estimation*:

   The final 10 diverging samples are used to test the predictive performance of the neutral class parameters. This assesses whether the neutral class captures behaviour in interactive yet non-threatening scenarios.

For each of the 20 samples drawn from both the free-flow and diverging interaction subsets, sample sizes are matched to that of the original interaction dataset to enable fair comparison. This approach is critical because imbalanced data can distort model evaluation results in terms of accuracy and interpretability. When subsets differ significantly in size, predictive errors may reflect differences in data volume rather than true behavioural deviations. This technique ensures that any observed differences in predictive performance are attributable to the behavioural context rather than sample size effects, allowing for a more robust and unbiased validation of the neutral class parameters.

Figure 4 presents a summary of the validation framework used to assess the predictive accuracy of the neutral class across different driving contexts. For each data subset, the distance error between observed choices (positions) and predicted choices is calculated using four prediction methods:

1. Arithmetic Mean: Expected position weighted by choice probabilities.

2. Harmonic Mean: Probability-weighted inverse average to emphasise high-probability outcomes.

3. Most Probable Alternative: Selecting the alternative with the highest probability.

4. Random Draw from CDF: Drawing an alternative based on the Cumulative Distribution Function (CDF) of probabilities.

The average, minimum, and maximum error are computed across the 10 test samples for each data subset and for each evaluation method.

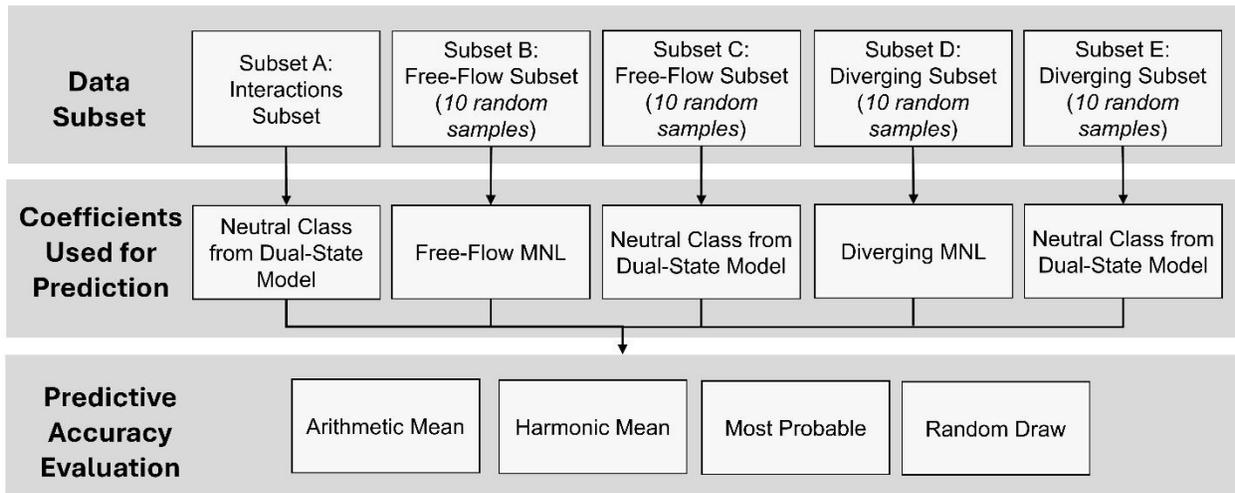

*Figure 4 Summary of the validation framework used to evaluate the predictive accuracy of the neutral class across multiple driving contexts*

## 4. Application to Proposed Framework

### 4.1. Dataset

The proposed framework is applied to the publicly available rounD dataset, which includes naturalistic video recordings from three German roundabouts (Krajewski et al., 2020). Since the Neuweiler site contains the most extensive set of recordings, the analysis focuses exclusively on that location. Although the dataset includes eight road-user classes (bicycles, buses, cars, trucks, trailers, vans, motorcycles, and pedestrians), the analysis targets interactions involving passenger cars with other motor vehicles (buses, trucks, trailers, vans, and motorcycles). Non-motorised road users are excluded, as their inclusion would introduce fundamentally different movement dynamics and risk perceptions, potentially confounding class-membership estimation and reducing the model's ability to isolate driving-state transitions.

The dataset comprises multiple car trajectories, each spanning a different duration depending on the video recording. The raw data from recording number 2 to 23 excluding bicyclists and pedestrians includes 4,982,020 observations, each row corresponding to 0.04 seconds. Before modelling, two data-cleaning steps are performed. First, a polygonal catchment area is defined around the central circulatory section and retain only a small number of trajectories on the entry and exit approaches, removing all others as well as any paths from channelized right-turn lanes. Only vehicles that enter, circulate within, and exit the roundabout remain (see Figure A.1 in the Appendix). After applying this filtering, the total filtered observations of cars and other vehicles are 2,967,616 which includes 9,799 passenger cars and 1,839 other types of vehicles. Second, to guard against video-tracking errors, collision events are detected systematically. This is achieved by identified the collided (intersected) vehicle's start and end timestamps and remove its trajectory along with those of any vehicles that interact with it during that time window (the excluded "current" vehicle is shown in red and interacting vehicles in blue in Figure A.2 in the Appendix). The total number of observations (cars and other vehicles trajectories) excluded is 13,477.

The vehicle trajectories are processed at a 1-second interval, corresponding to every 25 rows in the raw dataset (each row represents 0.04 seconds). This yields a total of 97,701 observed car trajectory points at a 1-second resolution. A total of 879 observations that fall outside the predefined spatial grid are first removed, followed by an additional 1,303 observations immediately adjacent in time to ensure spatial consistency. An additional 9,765 choices occurring at the very beginning or end of each trajectory are then excluded, as these lack either a preceding or subsequent frame required to compute changes in speed and direction. After applying these filters, 85,754 valid observations remain for analysis and for computing the conflict indicator. Table 2 summarises the full sequence of data filtering steps. The Spatio-Temporal Composite Proximity (STCP) indicator is utilized (Al-Haideri et al., 2025) is used to segment the data and is described in detail in the following section. Based on the computed STCP values, the 85,754 observations are categorised into three subsets for model estimation and validation: interactions data (8,813 observations), diverging interaction data (53,102 observations), and free-flow data (23,839 observations).

*Table 1 Summary of data filtering steps applied to passenger-car observations (rounD dataset)*

| *Cars' observations filtering* | *Removed* | *Remaining* |
|---|---|---|
| Initial observations | - | 97,701 |
| Excluded: outside spatial grid | 879 | 96,822 |
| Excluded: adjacent to invalid grid points | 1,303 | 95,519 |
| Excluded: first/last time step of each car trajectory | 9,765 | 85,754 |
| Final processed cars observations | - | 85,754 |

### 4.2. Model Specification

The MNL structure is adopted for both the class membership and conditional choice probabilities in the dual-state model. The MNL is widely used due to its computational efficiency and closed-form solution, making it especially suitable for large-scale trajectory datasets. The class membership component is specified using the Closing Time-to-Collision (CTTC) conflict indicator (Al-Haidari et al., 2025). CTTC estimates the time remaining before a potential collision, based on the relative velocity and position of the interacting vehicle, assuming constant velocity extrapolation. To ensure that it captures meaningful threat rather than incidental proximity, CTTC is only computed for converging vehicles—those on a path likely to result in interaction.

CTTC is used in the class membership model because it reflects a driver's general sense of situational risk rather than an immediate response to a specific event. As such, it serves as a proxy for how drivers internally perceive and evaluate ongoing traffic conditions, which aligns with the purpose of the class membership component: to estimate the likelihood of being in a defensive versus neutral state based on broader environmental awareness. Drivers are assumed to occupy one of two latent behavioural states at each time step $t$: defensive or neutral. The probability that driver $n$ is in the defensive state at time $t$ is:

$$P_{nt(Def)} = \frac{e^{W_{nt}^{Def}}}{1 + e^{W_{nt}^{Def}}}$$

For identification, the utility of the neutral state is normalized to zero:

$$W_{nt}^{Neut} = 0$$

The utility of the defensive state is defined as:

$$W_{nt}^{Def} = \alpha_{CTTC} \frac{1}{1 + CTTC_{nkt}^{min}} + \alpha_{ASC}$$

Where:

- $CTTC_{nkt}^{min}$ is the minimum CTTC between driver $i$ and any interacting vehicle $k$ at time $t$; and
- $\alpha_{CTTC}$ and $\alpha_{ASC}$ are to be estimated.

A driver-memory dummy was incorporated, defined as a lagged binary indicator equal to 1 if the driver's posterior probability of being in the defensive state was ≥ 0.5 at the previous time step. This addition aimed to capture behavioural persistence and stabilise the inferred sequence of states. However, the model failed to invert with this specification. As a result, choices made by the same driver are treated independently in the current model structure. The probability that driver $n$ selects alternative $i$, conditional on being in state $s$ is given by:

$$P_{nit}|s = \frac{e^{V_{nit}^s}}{\sum_J V_{mit}^s}$$

The systematic utility function $V_{nit}^s$ is specified differently depending on the behavioural state:

$$V_{nit}^s = \begin{cases} \beta_D I_{Dnit} STCP_{nKt} + \beta_T [I_{nit}^{right} CAI_{ikt}^{left} + I_{nit}^{left} CAI_{ikt}^{right}] + \beta_C I_{Cnit}, & s = Defensive \\ \gamma_D I_{Dnit} + \gamma_A I_{Anit} + \gamma_T I_{Tnit} + \gamma_C I_{Cnit} + \gamma_{Dist} Dist_{nit}, & s = Neutral \end{cases}$$

Where:

- $STCP_{nKt}$ is the Spatio-Temporal Composite Proximity perceived by driver $i$ from interacting vehicles $K$ during time $t$;
- $I_{nit}^{right}$ = 1 if the driver selects alternatives 3, 6, or 9 (left-turn alternatives), 0 otherwise;
- $I_{nit}^{left}$ = 1 if the driver selects alternatives 1, 4, or 7 (right-turn alternatives), 0 otherwise; and
- $CAI_{ikt}^{left}$ and $CAI_{ikt}^{right}$ are Collision Angle Intensities, which quantify the directional severity of a potential collision between driver $i$ and interacting vehicle $k$ at time $t$, based on their relative positions and movement directions. These intensities are calculated only for the interacting vehicle with the minimum CTTC, that is, the vehicle posing the most immediate threat. This ensures that the model activates directional evasive turning responses specifically in response to a highest risk interaction. The intensity values range from 0 (no directional threat) to 1 (maximum directional threat), and are defined as:
  - $CAI_{ikt}^{right} = |\sin(\alpha)|$, for collision angles $0° \leq \alpha \leq 180°$
  - $CAI_{ikt}^{left} = |\sin(\alpha)|$, for collision angles $180° \leq \alpha \leq 360°$.

The terms $\beta_D I_{Dnit} STCP_{nKt}$ and $\beta_T [I_{nit}^{right} CAI_{ikt}^{left} + I_{nit}^{left} CAI_{ikt}^{right}]$ represent the collision avoidance mechanism within the defensive state. The first term captures responsiveness to proximity-based risks, while the second represents evasive steering based on lateral threat direction.

The STCP indicator is used in the conditional choice probability and offers a high-resolution measure of perceived risk at the individual level. Unlike CTTC, STCP is designed to capture evasive response potential, incorporating both spatial closeness and temporal imminence. It does not require road users to be on a direct collision course. Instead, it reflects likely evasive actions based on nearby movement dynamics. The STCP for driver $n$ interacting with vehicles $K$ during time $t$ is expressed as:

$$STCP_{nKt} = \sum_{k=1}^{K} \frac{1}{(1 + MG_{nkt})(1 + TTMG_{nkt})}$$

Where $MG_{nkt}$ represents the minimum distance between car $i$ and interacting vehicle $k$ at time instant $t$, assuming a constant velocity extrapolation, and $TTMG_{nkt}$ is the time that interacting vehicle $k$ would take to reach that minimum distance.

Figures 5 and 6 illustrate the behavioural relevance of the STCP indicator. To assess whether the indicator supports a dual-state behavioural interpretation, k-means clustering was applied with k = 2 to the 8,813 STCP values. As shown in Figure 5, this yields two well-separated clusters, consistent with the hypothesised segmentation into low-risk (neutral) and high-risk (defensive) states. Figure 6 visualises the STCP surface and demonstrates how it peaks under close, imminent threats (i.e., small MG and TTMG) and declines as either component increases. It is important to note that the spatial (MG) and temporal (TTMG) components of the STCP indicator are not normalised or scaled in the current formulation. This means that values such as 1 m and 1 s result in an STCP of 0.25, which mathematically suggests a low-risk scenario (closer to 0), despite being behaviourally indicative of high-risk conditions. Future work should explore appropriate scaling or transformation of these components to better align the numerical output of the indicator with real-world perceptions of risk.

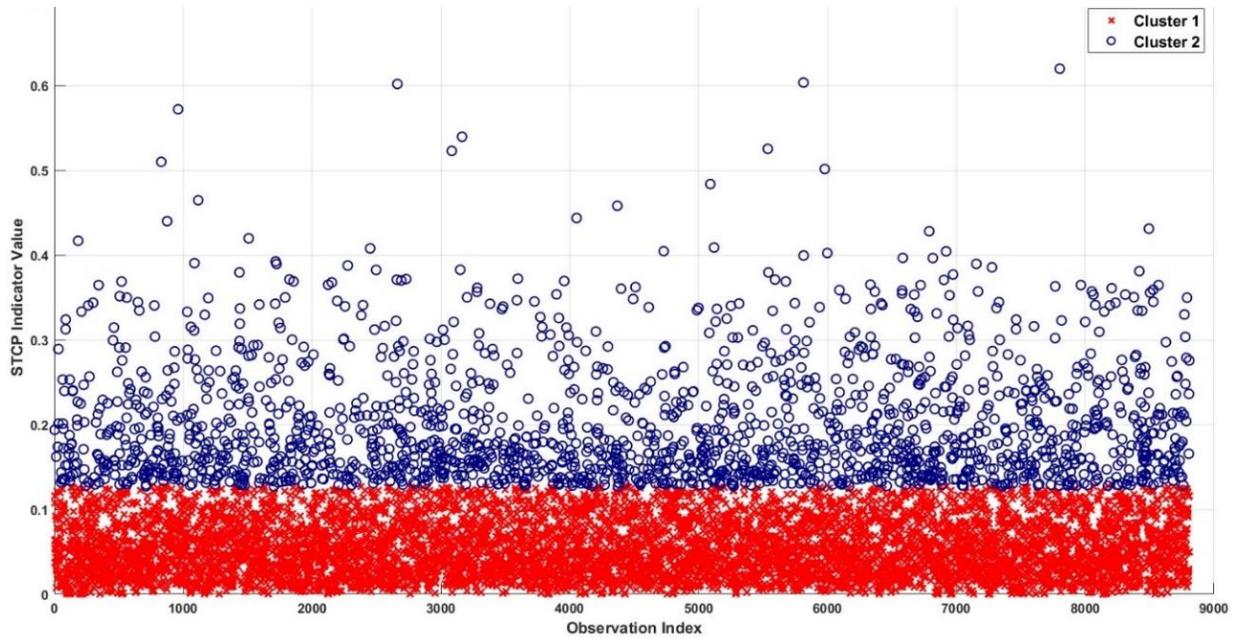
*Figure 5 k-means clustering for the STCP indicator*

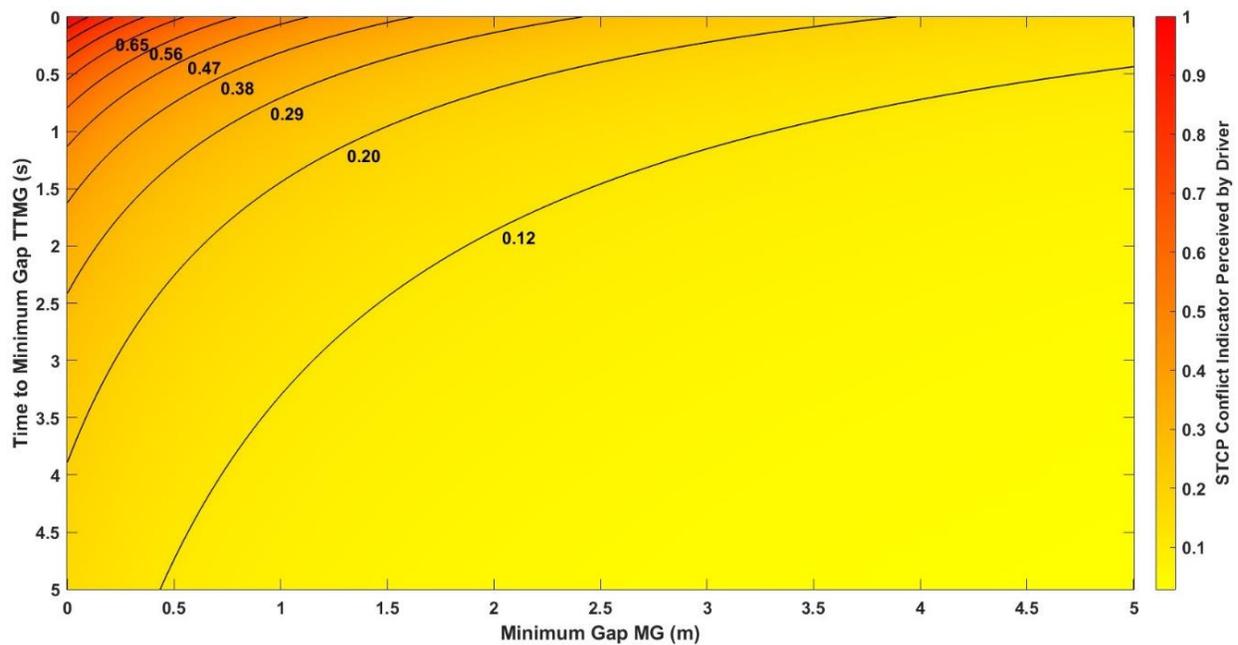
*Figure 6 STCP perceived by a driver from an interacting vehicle*

### 4.3. Estimation Results and Discussions

Data processing is performed in MATLAB, while model estimation is carried out in GAUSS 25 using an EM algorithm coupled with MLE (Aptech Systems, 2025). As an initial demonstration that a latent structure provides a better fit than a single-state model, two specifications using the Interactions Data subset are estimated: (1) a standard MNL model with seven parameters, and (2) the dual-state latent DCM model representing defensive versus neutral driving states, with a total of twelve parameters. Table 4 summarises the model fit statistics for both specifications.

*Table 2 Fit statistics for the MNL model versus latent DCM*

| Model (interactions subset) | df | Mean log-likelihood | AIC | BIC |
|---|---|---|---|---|
| MNL | 7 | -1.72314 | 30,386.07 | 30,435.65 |
| Dual-State Latent DCM | 12 | -1.58681 | 27,989.11 | 28,078.12 |

As shown in Table 4, the dual-state latent DCM delivers a clearly superior fit compared to the single-state MNL. Its mean log-likelihood of –1.58681 is substantially higher than the MNL's –1.72314, despite estimating five additional parameters. This improvement is reflected in the information criteria: the dual-state's Akaike Information Criterion (AIC) of 27,989.11 and Bayesian Information Criterion (BIC) of 28,078.12 are each more than 2,000 points lower than those of the MNL (AIC = 30,386.07; BIC = 30,435.65). These results indicate that the latent DCM provides a significantly better fit without evidence of overfitting, and that the gain in explanatory power justifies the modest increase in model complexity. Overall, the findings support the effectiveness of explicitly modelling dual behavioural states over a conventional single-state logit in representing driver decision-making at roundabouts.

Table 5 presents the dual-state latent DCM estimates using the interactions subset. All parameters are statistically significant at the 5% confidence level At the class-membership level, the CTTC coefficient in Class 1 is large and positive, indicating that when another vehicle is approaching the driver (*i.e.* closing in), the likelihood of becoming defensive increases sharply. The negative coefficient for the ASC sets the neutral state as the default state in the absence of conflict. Since the utility of the neutral state is normalised to zero, it serves as the reference class. In other words, when the CTTC does not exert a strong positive influence, drivers default to the neutral state. This interpretation indicates that Class 1 captures the defensive state, while Class 2 reflects routine, neutral driving.

At the conditional choice level, the defensive state (Class 1) is characterised by heightened responsiveness to spatial and temporal proximity, as captured by the deceleration STCP variable, which yields a large positive coefficient. This indicates that drivers are strongly inclined to decelerate when nearby vehicles pose a potential threat, *i.e.*, when one or more interacting vehicles are close in both space and time, reflecting a clear collision-avoidance mechanism. The turning with CAI variable also has a large positive coefficient, suggesting that drivers prefer to turn as an evasive manoeuvre when a lateral threat is present, as inferred from the CAI on either the right or left. The coefficient of the choice availability variable is positive in both behavioural classes but is notably larger in the defensive class. This suggests that, although all drivers tend to favour accessible options, those in the defensive state are more sensitive to the availability of safe, geometrically viable alternatives and are more likely to avoid options that appear blocked or constrained.

In the neutral state (Class 2), the estimated coefficients reflect baseline driving tendencies under low-risk conditions. The positive coefficient for deceleration dummy suggests a moderate but general preference for slowing down, likely reflecting routine speed adjustments rather than urgent avoidance. The positive coefficient for the acceleration dummy indicates that drivers in the neutral state also favour speeding up when appropriate, highlighting a balanced responsiveness to traffic flow. The coefficient for the turning dummy is also positive, suggesting that turning manoeuvres are commonly undertaken in the neutral state likely to maintain their lane and path alignment. These positive values for acceleration, deceleration and turning reinforce that the neutral class does not correspond to passive or unengaged behaviour but rather to typical, context-appropriate actions. The negative coefficient for distance to the ideal path variable implies that drivers in this state tend to favour options that keep them aligned with their ideal path through the roundabout.

To evaluate the impact of how conflict indicators are integrated into the model, an alternative specification was estimated in which STCP was placed in the class membership model and CTTC in the conditional choice probabilities. This structure, pairing a highly specific, short-range indicator with the broader behavioural state and using a general risk measure for immediate decisions, resulted in a significantly worse model fit. The performance gap reinforces the rationale for the original formulation, where broader indicators inform the driver's behavioural state and high-resolution indicators guide moment-to-moment choices. Estimation results of this alternative model are provided in Table A.1 in the Appendix.

*Table 3 Dual-state latent DCM model estimation results using interactions subset*

| Class membership probability ||||||||
|---|---|---|---|---|---|---|---|
| *Class 1 (Defensive state)* |||| *Class 2 (Neutral state)* ||||
| Coefficient | Estimate | Standard error | $t$-stat | | | | |
| $\alpha_{CTTC}$ (Closing Time-to-Collision) | 13.7651 | 0.5778 | 23.824* | 0 (reference class) |||| 
| $\alpha_{ASC}$ (Alternative-Specific Constant) | -3.5432 | 0.1726 | -20.53* | |||| 
| Average Aggregate Class Probability | 43.5% ||| 56.5% ||||
| Class-specific choice probability ||||||||
| *Class 1 (Defensive state)* |||| *Class 2 (Neutral state)* ||||
| Coefficient | Estimate | Standard error | $t$-stat | Coefficient | Estimate | Standard error | $t$-stat |
| $\beta_D$ (Deceleration STCP) | 9.3995 | 0.6151 | 15.282* | $\gamma_D$ (Deceleration) | 1.2916 | 0.0435 | 29.663* |
| $\beta_C$ (Choice Availability) | 1.1381 | 0.1208 | 9.419* | $\gamma_C$ (Choice Availability) | 0.7226 | 0.0726 | 9.947* |
| $\beta_T$ (Turning with CAI) | 9.2496 | 0.9113 | 10.15* | $\gamma_T$ (Turning) | 0.9578 | 0.0387 | 24.745* |
| | | | | $\gamma_A$ (Acceleration) | 0.9421 | 0.0601 | 15.678* |
| | | | | $\gamma_{Dist}$ (Distance to Ideal Path) | -0.6081 | 0.0213 | -28.582* |
| Mean log-likelihood at convergence | -1.58681 |||||||
| Number of observations | 8,813 (interactions subset) |||||||

*Statistically significant at the 5% confidence level

Figure 7 depicts the relationship between CTTC and the probability of a driver being in the defensive state. The graph is plotted only until the probability of being in the neutral state reaches 90%, as the curve becomes flat beyond that point and is unlikely to offer additional behavioural insights. As illustrated, the probability of being in the defensive state is highest at very low CTTC values and gradually declines as CTTC increases, eventually levelling off. Several preliminary insights can be drawn from this figure. For example, a defensive probability of approximately 0.5 appears to correspond to CTTC values around 3 seconds. This value may warrant further investigation to determine whether it represents a meaningful behavioural threshold. Defensive probabilities above 0.5 could carry added behavioural significance, possibly marking transitions to different perceived severity levels, though these interpretations remain exploratory.

The diagonal line in the figure serves as a reference, indicating where CTTC would scale linearly with class membership probability. This benchmark helps identify which observations fall above or below it, suggesting a potential non-linear relationship between CTTC and defensive state membership. Notably, the shape of this non-linearity is not predefined, it emerges from the data. This data-driven feature of the model enables behavioural distinctions to surface naturally, without imposing a predetermined functional form. While this figure provides useful insight into how perceived time gaps may relate to behavioural states, these interpretations remain provisional. It is uncertain whether the thresholds observed here are generalisable or whether similar patterns would appear in other datasets. Further investigation is needed to validate the robustness and transferability of these findings.

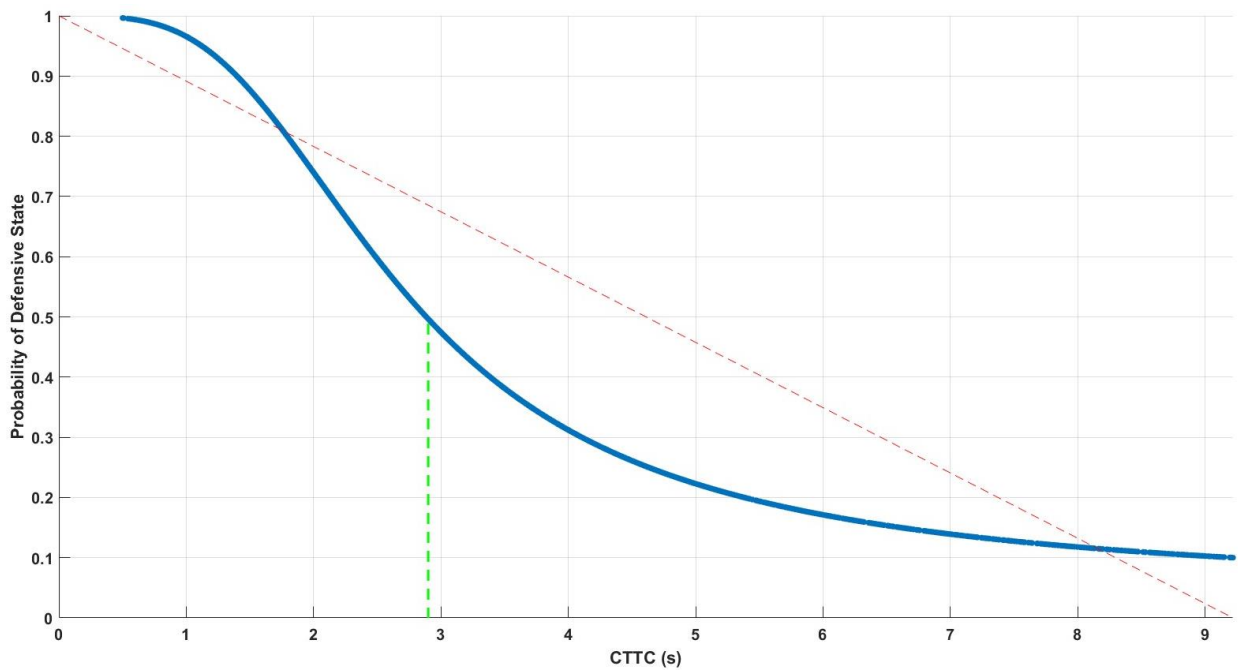

*Figure 7 Relationship between Closing Time to Collision (CTTC) and the probability of being in a defensive state*

### 4.4. Validation of the Neutral State

Table 6 presents the average distance error (in metres) between predicted and observed vehicle positions across the five data subsets and four prediction methods. These results provide insight into the predictive accuracy, stability, and potential generalisability of the neutral class parameters derived from the dual-state model. The similarity matrices for the 20 random samples drawn of each subset from the free-flow and diverging are presented in Tables A.3-A.6 in the Appendix.

As expected, Subset A, where the neutral class coefficients are reapplied to the same interaction data it was estimated from, yields the lowest average prediction errors across all methods. The most probable method achieves an average error of 4.247 m, establishing a benchmark for internal consistency and demonstrating that the model effectively recovers in-sample behaviour.

When the neutral class coefficients are applied to the out-of-sample data, as in Subsets C (free-flow) and E (diverging), average errors increase slightly but remain within a comparable range. For example, in Subset E, the neutral class achieves a most probable error of 4.997 m, only marginally higher than the diverging-specific MNL model's 4.844 m using Subset D. This indicates that the neutral state maintains reasonable predictive power in moderate, low-threat driving scenarios, supporting its relevance beyond its original estimation context.

The practical significance of these error differences remains an open question. While there is currently no empirical benchmark to define what constitutes an "acceptable" prediction error in behavioural choice contexts, the results offer promising preliminary evidence. The difference in average errors between Subset A and other subsets hovers around 0.5 metres across methods. Though seemingly modest, such differences may reflect meaningful shifts in prediction performance, suggesting that the model does extract a behaviourally coherent neutral state. This is interpreted as indicative of some degree of success in isolating a transferable behavioural mode, but further work is needed to contextualise these values against practical thresholds or real-world applications.

Moreover, stability across repeated random samples is particularly noteworthy. In Subset B, the most probable method using the free-flow MNL model yields a prediction error range of 0.038 m. Whereas in Subset C, where the neutral class is applied to similar free-flow conditions, the corresponding range narrows to just 0.021 m. This reduced variation supports the notion that the neutral class produces more stable and generalisable predictions. It also suggests that the neutral class, estimated on the interactions data, avoids potential overfitting issues that may arise when training and testing on the same data domain, as in the free-flow MNL case.

Overall, while additional research is needed to establish formal diagnostic standards for acceptable behavioural prediction error, the observed differences in both average accuracy and prediction stability provide early evidence for the promise of the neutral class as a transferable, low-conflict behavioural profile. Future work should explicitly examine the practical implications of these modest prediction differences, particularly in safety-critical or simulation-based applications where small variations in position may influence decision outcomes. Prediction accuracy results of the alternate model are provided in Table A.2 in the Appendix.

*Table 4 Average distance error in meters between predicted and observed positions using different prediction methods*

| Data subset | Coefficients used for prediction | Arithmetic mean | | | Harmonic mean | | | Most probable | | | Random draw (CDF) | | | Sample Size |
|---|---|---|---|---|---|---|---|---|---|---|---|---|---|---|
| Subset A: full interactions subset | Neutral Class from Latent DCM | 5.532 | | | 5.503 | | | 4.247 | | | 5.588 | | | |
| | | *Min. (10 random sample draws)* | *Avg. (10 random sample draws)* | *Max. (10 random sample draws)* | *Min. (10 random sample draws)* | *Avg. (10 random sample draws)* | *Max. (10 random sample draws)* | *Min. (10 random sample draws)* | *Avg. (10 random sample draws)* | *Max. (10 random sample draws)* | *Min. (10 random sample draws)* | *Avg. (10 random sample draws)* | *Max. (10 random sample draws)* | |
| Subset B: free-flow (first 10 samples) | Free-flow MNL | 6.044 | 6.062 | 6.088 | 6.050 | 6.069 | 6.096 | 5.108 | 5.121 | 5.146 | 6.116 | 6.137 | 6.169 | *8,813* |
| Subset C: free-flow (second 10 samples) | Neutral class from latent DCM | 6.806 | 6.821 | 6.837 | 6.816 | 6.832 | 6.846 | 5.321 | 5.333 | 5.342 | 6.878 | 6.893 | 6.907 | |
| Subset D: diverging interactions (first 10 samples) | Diverging MNL | 6.206 | 6.239 | 6.259 | 6.213 | 6.247 | 6.268 | 4.829 | 4.844 | 4.859 | 6.272 | 6.303 | 6.335 | |
| Subset E: diverging interactions (second 10 samples) | Neutral class from latent class model | 6.429 | 6.450 | 6.463 | 6.438 | 6.459 | 6.470 | 4.989 | 4.997 | 5.004 | 6.491 | 6.520 | 6.544 | |

## 5. Conclusions

This paper presents a behavioural framework that introduces a fourth dimension of traffic conflict severity: *the probability of being in a defensive state*, conceptualised as a latent, safety-relevant response to perceived risk. Building on established dimensions, proximity, consequence, and evasive action, this behavioural dimension reflects internal driver states that are not directly observable but are inferred from naturalistic trajectory data. Unlike conventional severity indicators, this probabilistic measure captures how drivers cognitively and affectively interpret external risk cues.

Driving is modelled as a probabilistic blend of two latent behavioural states: defensive and neutral. This framework is grounded in psychological theory, including the ABC triad and dual-process models of cognition, and it avoids rigid classification by estimating a continuous likelihood of state membership. This formulation acknowledges the fluid and context-dependent nature of behaviour, where transitions between states unfold along a continuum rather than at binary thresholds.

Rather than viewing behaviour as a snapshot, the model treats trajectories as behavioural traces, dynamic sequences that reflect how drivers adjust their decisions over time in response to their environment. To account for the spatial context of these decisions, the framework incorporates the concept of directional risk fields, modelled using a spatial grid informed by each driver's speed and heading. This allows for location-sensitive estimation of when and where defensive responses are likely to emerge.

By integrating psychological, temporal, and spatial dimensions, this framework provides a more nuanced and flexible approach to assessing road user behaviour. While the model offers a promising way to capture perceived risk and latent defensive behaviour, further work is required to validate its relationship with actual crash outcomes. Future research should also explore how this behavioural dimension interacts with other severity indicators across diverse traffic environments to ensure its robustness and generalisability.

The proposed framework is applied to naturalistic vehicle trajectories at a roundabout. In this application, CTTC conflict indicator is used in the class membership probability to reflect broader anticipatory risk, while the STCP indicator provided a more detailed lens on immediate spatial-temporal conflicts within the conditional choice probability. Estimation results suggest meaningful distinctions between the two latent states: drivers in the defensive state showed greater responsiveness to spatial and temporal threats, while those in the neutral state exhibited more balanced patterns of movement under lower-risk conditions. Although the framework currently excludes trajectories that include collisions due to tracking noise, it remains conceptually well-suited to capture such events, particularly at the extreme end of the defensive spectrum where CTTC approaches zero.

Validation results indicate that the neutral state performs reasonably well across both in-sample and out-of-sample conditions. Prediction errors remain low and stable, with only marginal increases when applied to new data subsets, suggesting that the estimated parameters could capture a transferable behavioural pattern under low-conflict conditions. While the practical significance of the observed differences remains unclear, the consistency in accuracy and limited variation across random samples point to the model's potential for generalisability. Further work is needed to evaluate how these results translate into real-world applications.

An area of particular interest is the framework's ability to highlight mismatches between inferred internal states and external risk indicators. For instance, a low defensive probability in a high-risk situation might reflect delayed awareness or misperception, while a high defensive probability in a low-risk setting could suggest over-caution or elevated sensitivity. These mismatches may offer an additional layer of insight into latent vulnerabilities, though further investigation is needed to interpret them reliably.

Several limitations should be acknowledged. Future work should explore scaling or normalising the spatial (MG) and temporal (TTMG) components within the STCP indicator, which are currently treated in raw units. Without accounting for perceptual thresholds or behavioural sensitivity, the indicator may underestimate high-risk conditions, for example, when both spatial and temporal gaps are narrow but not extreme. The neutral state may also involve unmodelled behavioural variation, such as distinctions between free-flow and diverging manoeuvres, which are currently treated uniformly. Incorporating additional latent classes could help capture this heterogeneity. Finally, the assumption of independence across drivers may restrict the model's ability to represent behavioural persistence or memory effects within individual driving patterns.

# Appendix

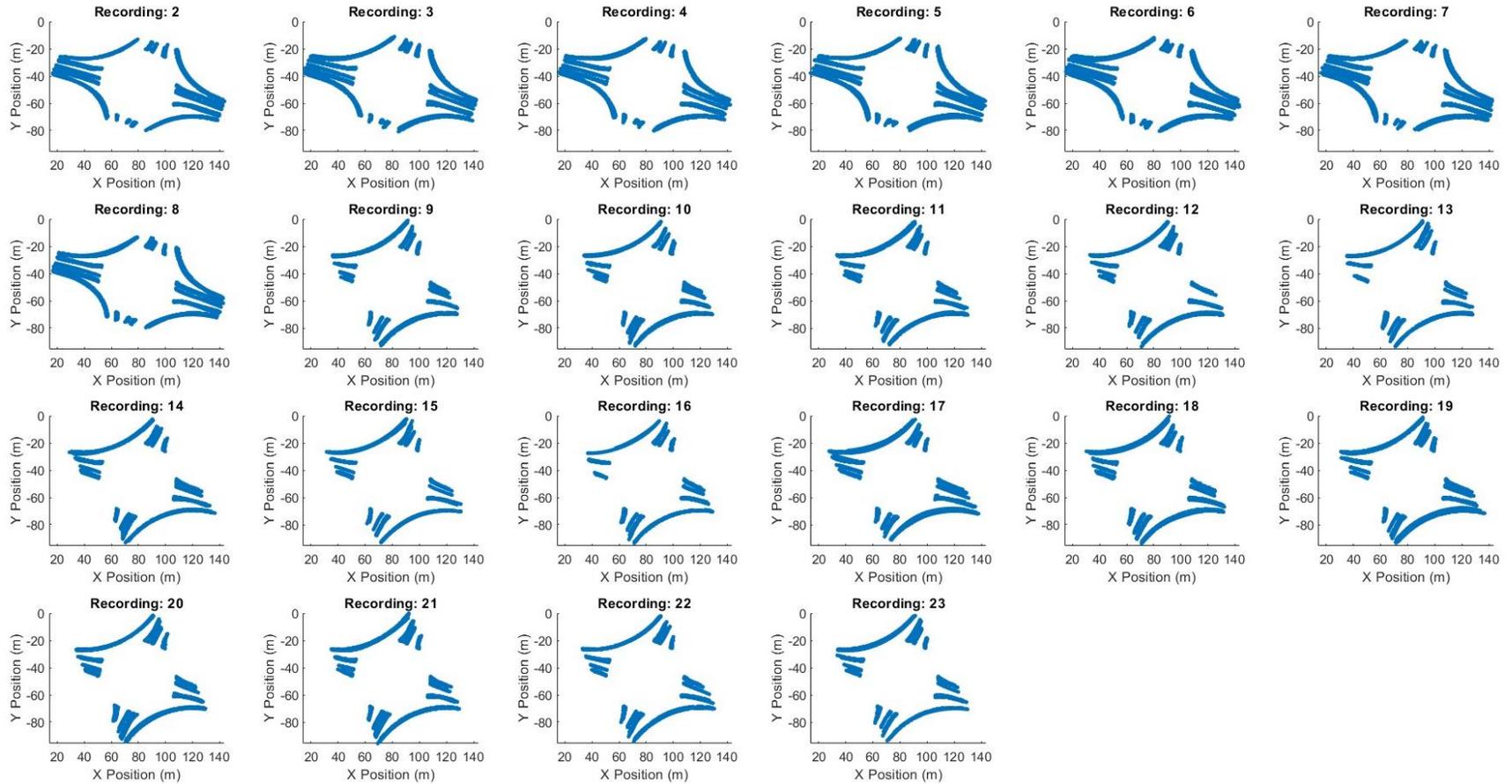

*Figure A.1 Trajectories of raw vehicles excluded within the polygonal catchment area defined around the central circulatory section of the roundabout.*

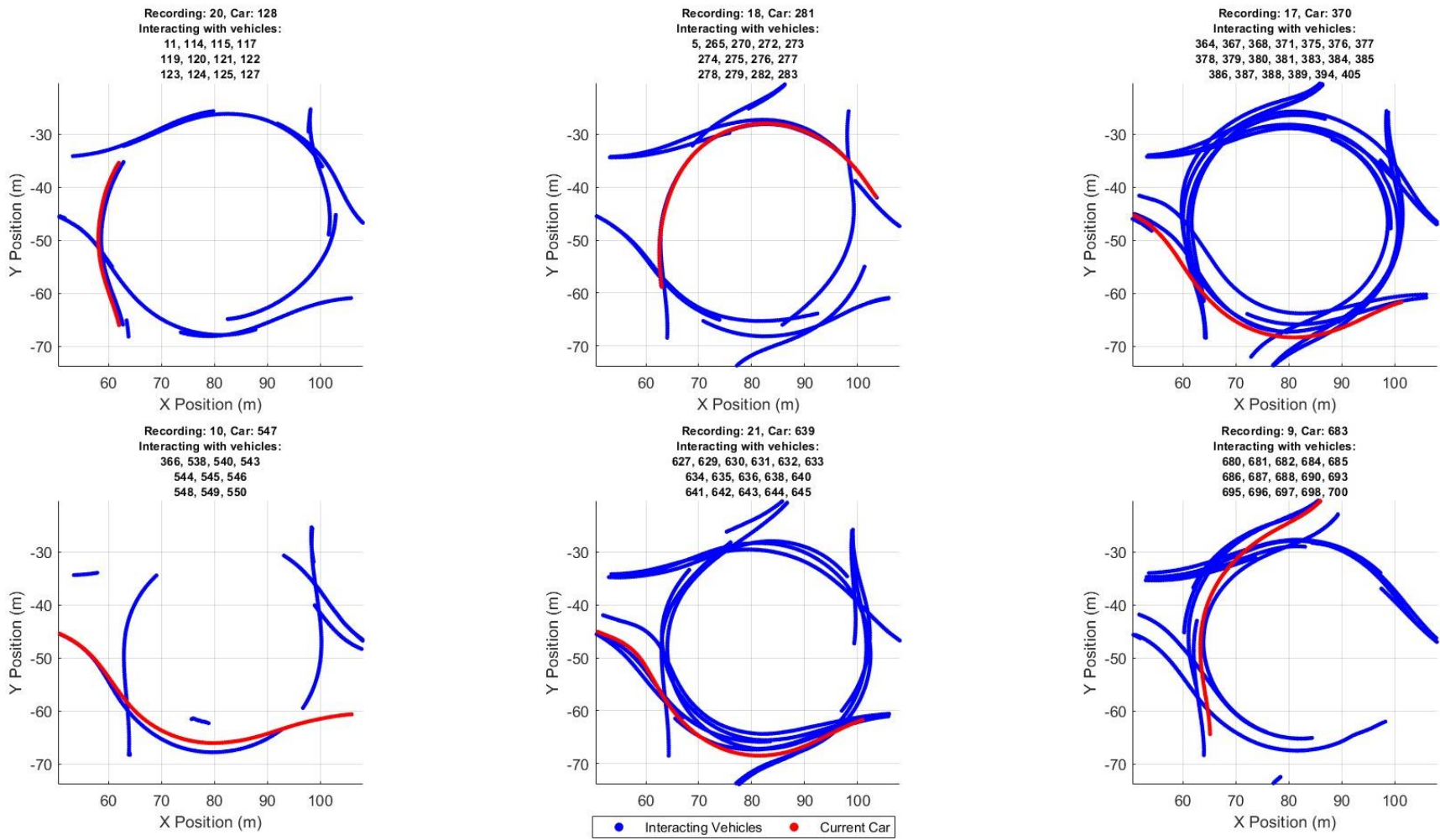

*Figure A.2 Trajectories of the car being analysed that is involved in a detected collision (red trajectories) and the trajectories of any interacting vehicles during the same time window (blue trajectories)*

*Table A.1* ***Alternate*** *latent DCM alternate model estimation results for dual-state driver behaviour using interactions subset*

| | Class membership probability | | | | | | |
|---|---|---|---|---|---|---|---|
| | *Class 1 (Defensive state)* | | | *Class 2 (Neutral state)* | | | |
| Coefficient | Estimate | Standard error | $t$-stat | | | | |
| $\alpha_{STCP}$ (STCP) | 11.2296 | 0.8163 | 13.757 | 0 (reference class) | | | |
| $\alpha_{ASC}$ (Alternative-Specific Constant) | -2.9834 | 0.1171 | -25.483 | | | | |
| | Class-specific choice probability | | | | | | |
| | *Class 1 (Defensive state)* | | | *Class 2 (Neutral state)* | | | |
| Coefficient | Estimate | Standard error | $t$-stat | Coefficient | Estimate | Standard error | $t$-stat |
| $\beta_D$ (Deceleration CTTC) | 4.214 | 0.4948 | 8.517 | $\gamma_D$ (Deceleration) | 1.6175 | 0.0442 | 36.568 |
| $\beta_C$ (Choice Availability) | 1.6207 | 0.2457 | 6.596 | $\gamma_C$ (Choice Availability) | 1.0114 | 0.0563 | 17.979 |
| $\beta_T$ (Turning) | -0.5257 | 0.1017 | -5.168 | $\gamma_T$ (Turning) | 2.4447 | 0.0927 | 26.379 |
| | | | | $\gamma_A$ (Acceleration) | 1.6009 | 0.0485 | 33.013 |
| | | | | $\gamma_{Dist}$ (Distance to Ideal Path) | -0.6798 | 0.0193 | -35.311 |
| Mean log-likelihood at convergence | -1.71426 | | | | | | |
| Number of observations | 8,813 (interactions subset) | | | | | | |

Table A.2 Average distance error in meters between predicted and observed positions for the **alternate** model

| Data subset | Coefficients used for prediction | Arithmetic mean | | | Harmonic mean | | | Most probable | | | Random draw (CDF) | | | Sample Size |
|---|---|---|---|---|---|---|---|---|---|---|---|---|---|---|
| Interactions subset | Neutral Class from DCM Model {alternate model} | 5.644 | | | 5.611 | | | 4.803 | | | 5.711 | | | |
| | | Avg. (10 random sample draws) | Min. (10 random sample draws) | Max. (10 random sample draws) | Avg. (10 random sample draws) | Min. (10 random sample draws) | Max. (10 random sample draws) | Avg. (10 random sample draws) | Min. (10 random sample draws) | Max. (10 random sample draws) | Avg. (10 random sample draws) | Min. (10 random sample draws) | Max. (10 random sample draws) | *8,813* |
| Subset C: free-flow (second 10 samples) | Neutral Class from Latent DCM {alternate model} | 6.974 | 6.957 | 6.993 | 6.984 | 6.965 | 7.004 | 6.002 | 5.960 | 6.029 | 7.054 | 7.024 | 7.071 | |
| Subset E: diverging interactions (second 10 samples) | Neutral Class from Latent DCM {alternate model} | 6.597 | 6.572 | 6.629 | 6.605 | 6.579 | 6.637 | 5.627 | 5.594 | 5.667 | 6.678 | 6.647 | 6.702 | |

*Table A.3 Free-flow subset first 10 samples similarity*

| Sample | Sample 1 | Sample 2 | Sample 3 | Sample 4 | Sample 5 | Sample 6 | Sample 7 | Sample 8 | Sample 9 | Sample 10 |
|---|---|---|---|---|---|---|---|---|---|---|
| Sample 1 | 8813 | 3274 | 3324 | 3289 | 3194 | 3176 | 3275 | 3190 | 3283 | 3284 |
| Sample 2 | 3274 | 8813 | 3257 | 3282 | 3215 | 3257 | 3282 | 3290 | 3278 | 3301 |
| Sample 3 | 3324 | 3257 | 8813 | 3337 | 3293 | 3248 | 3223 | 3258 | 3221 | 3294 |
| Sample 4 | 3289 | 3282 | 3337 | 8813 | 3276 | 3225 | 3290 | 3228 | 3311 | 3228 |
| Sample 5 | 3194 | 3215 | 3293 | 3276 | 8813 | 3291 | 3257 | 3287 | 3283 | 3191 |
| Sample 6 | 3176 | 3257 | 3248 | 3225 | 3291 | 8813 | 3248 | 3258 | 3243 | 3205 |
| Sample 7 | 3275 | 3282 | 3223 | 3290 | 3257 | 3248 | 8813 | 3261 | 3250 | 3234 |
| Sample 8 | 3190 | 3290 | 3258 | 3228 | 3287 | 3258 | 3261 | 8813 | 3260 | 3240 |
| Sample 9 | 3283 | 3278 | 3221 | 3311 | 3283 | 3243 | 3250 | 3260 | 8813 | 3262 |
| Sample 10 | 3284 | 3301 | 3294 | 3228 | 3191 | 3205 | 3234 | 3240 | 3262 | 8813 |

*Table A.4 Free-flow subset second 10 samples similarity*

| Sample | Sample 1 | Sample 2 | Sample 3 | Sample 4 | Sample 5 | Sample 6 | Sample 7 | Sample 8 | Sample 9 | Sample 10 |
|---|---|---|---|---|---|---|---|---|---|---|
| Sample 1 | 8813 | 3254 | 3211 | 3332 | 3301 | 3170 | 3251 | 3308 | 3204 | 3201 |
| Sample 2 | 3254 | 8813 | 3214 | 3331 | 3304 | 3206 | 3279 | 3252 | 3200 | 3234 |
| Sample 3 | 3211 | 3214 | 8813 | 3299 | 3296 | 3233 | 3239 | 3290 | 3328 | 3294 |
| Sample 4 | 3332 | 3331 | 3299 | 8813 | 3235 | 3232 | 3178 | 3285 | 3306 | 3255 |
| Sample 5 | 3301 | 3304 | 3296 | 3235 | 8813 | 3274 | 3283 | 3225 | 3257 | 3282 |
| Sample 6 | 3170 | 3206 | 3233 | 3232 | 3274 | 8813 | 3285 | 3265 | 3193 | 3282 |
| Sample 7 | 3251 | 3279 | 3239 | 3178 | 3283 | 3285 | 8813 | 3195 | 3280 | 3280 |
| Sample 8 | 3308 | 3252 | 3290 | 3285 | 3225 | 3265 | 3195 | 8813 | 3298 | 3257 |
| Sample 9 | 3204 | 3200 | 3328 | 3306 | 3257 | 3193 | 3280 | 3298 | 8813 | 3243 |
| Sample 10 | 3201 | 3234 | 3294 | 3255 | 3282 | 3282 | 3280 | 3257 | 3243 | 8813 |

*Table A.5 Diverging subset first 10 samples similarity table*

| Sample | Sample 1 | Sample 2 | Sample 3 | Sample 4 | Sample 5 | Sample 6 | Sample 7 | Sample 8 | Sample 9 | Sample 10 |
|---|---|---|---|---|---|---|---|---|---|---|
| Sample 1 | 8813 | 1452 | 1460 | 1494 | 1464 | 1441 | 1459 | 1515 | 1459 | 1403 |
| Sample 2 | 1452 | 8813 | 1447 | 1479 | 1467 | 1506 | 1537 | 1498 | 1460 | 1451 |
| Sample 3 | 1460 | 1447 | 8813 | 1462 | 1489 | 1495 | 1427 | 1453 | 1455 | 1480 |
| Sample 4 | 1494 | 1479 | 1462 | 8813 | 1447 | 1493 | 1466 | 1483 | 1438 | 1494 |
| Sample 5 | 1464 | 1467 | 1489 | 1447 | 8813 | 1486 | 1496 | 1488 | 1465 | 1489 |
| Sample 6 | 1441 | 1506 | 1495 | 1493 | 1486 | 8813 | 1481 | 1510 | 1460 | 1449 |
| Sample 7 | 1459 | 1537 | 1427 | 1466 | 1496 | 1481 | 8813 | 1464 | 1490 | 1515 |
| Sample 8 | 1515 | 1498 | 1453 | 1483 | 1488 | 1510 | 1464 | 8813 | 1503 | 1462 |
| Sample 9 | 1459 | 1460 | 1455 | 1438 | 1465 | 1460 | 1490 | 1503 | 8813 | 1464 |
| Sample 10 | 1403 | 1451 | 1480 | 1494 | 1489 | 1449 | 1515 | 1462 | 1464 | 8813 |

*Table A.6 Diverging subset second 10 samples similarity table*

| Sample | Sample 1 | Sample 2 | Sample 3 | Sample 4 | Sample 5 | Sample 6 | Sample 7 | Sample 8 | Sample 9 | Sample 10 |
|---|---|---|---|---|---|---|---|---|---|---|
| Sample 1 | 8813 | 1428 | 1489 | 1451 | 1465 | 1436 | 1466 | 1490 | 1503 | 1487 |
| Sample 2 | 1428 | 8813 | 1461 | 1452 | 1437 | 1465 | 1490 | 1420 | 1470 | 1468 |
| Sample 3 | 1489 | 1461 | 8813 | 1493 | 1428 | 1466 | 1489 | 1506 | 1455 | 1454 |
| Sample 4 | 1451 | 1452 | 1493 | 8813 | 1445 | 1460 | 1428 | 1420 | 1422 | 1427 |
| Sample 5 | 1465 | 1437 | 1428 | 1445 | 8813 | 1486 | 1488 | 1484 | 1535 | 1485 |
| Sample 6 | 1436 | 1465 | 1466 | 1460 | 1486 | 8813 | 1458 | 1405 | 1480 | 1480 |
| Sample 7 | 1466 | 1490 | 1489 | 1428 | 1488 | 1458 | 8813 | 1476 | 1399 | 1440 |
| Sample 8 | 1490 | 1420 | 1506 | 1420 | 1484 | 1405 | 1476 | 8813 | 1411 | 1462 |
| Sample 9 | 1503 | 1470 | 1455 | 1422 | 1535 | 1480 | 1399 | 1411 | 8813 | 1427 |
| Sample 10 | 1487 | 1468 | 1454 | 1427 | 1485 | 1480 | 1440 | 1462 | 1427 | 8813 |